\documentclass{emulateapj}
\usepackage{amssymb}
\usepackage{lscape}
\newcommand{\msun}{\,{\rm M_\odot}}
\newcommand{\Ha}{\,{\rm H\alpha}}

\newcommand{\kms}{km~s$^{-1}\,$}
\newcommand{\lya}{\,{\rm Ly\alpha}}

\begin{document}

\title{THE KINEMATICS OF MORPHOLOGICALLY SELECTED $z\sim2$ GALAXIES \\
  IN THE GOODS-NORTH FIELD\altaffilmark{1}}
\author{\sc Dawn K. Erb, Charles C. Steidel}
\affil{California Institute of Technology, MS 105--24, Pasadena, CA 91125} 
\author{\sc Alice E. Shapley}
\affil{Department of Astronomy, 601 Campbell Hall, University of
  California at Berkeley, Berkeley, CA 94720}  
\author{\sc Max Pettini}
\affil{Institute of Astronomy, Madingley Road, Cambridge CB3 0HA, UK}
\and
\author{\sc Kurt L. Adelberger}
\affil{Carnegie Observatories, 813 Santa Barbara Street, Pasadena, CA 91101} 
\email{dke@astro.caltech.edu}

\submitted{Accepted for publication in ApJ}

\shorttitle{KINEMATICS OF $z\sim2$ GALAXIES}
\shortauthors{ERB ET AL.}

\altaffiltext{1}{Based on data obtained at the 
W.M. Keck Observatory, which is operated as a scientific partnership
among the California Institute of Technology, the University of
California, and NASA, and was made possible by the generous financial
support of the W.M. Keck Foundation.}

\begin{abstract}

We present near-IR spectra of $\Ha$ emission from 13 galaxies at
$z\sim2$ in the GOODS-N field.  The galaxies were selected primarily because
they appear to have elongated morphologies, and slits were aligned with
the major axes (as determined from the rest-frame UV emission) of 11 of
the 13.  If the galaxies are elongated because they are highly
inclined, alignment of the slit and major axis should maximize the
observed velocity and reveal velocity shear, if present. In spite of
this alignment, we see spatially resolved velocity shear in only two
galaxies.  We 
show that the seeing makes a large difference in the observed velocity
spread of a tilted emission line, and use this information to place
limits on the velocity spread of the ionized gas of the galaxies in
the sample: we 
find that all 13 have $v_{0.5} \leq 110$ \kms, where $v_{0.5}$ is the
velocity shear (half of the velocity range of a tilted emission
line) that would be observed under our best seeing conditions of
$\sim0\farcs5$.  When combined with previous work, our data also  
indicate that aligning the slit along the major axis does not
increase the probability of observing a tilted emission line.  We then
focus on the one-dimensional velocity dispersion $\sigma$, which is much less
affected by the seeing, and see that the elongated subsample exhibits
a significantly 
{\it lower} velocity dispersion than galaxies selected at random from
our total $\Ha$ sample, not higher as one might have expected.
We also see some evidence that the elongated galaxies are less
reddened than those randomly selected using only UV colors.  Both of
these results are counter to what would be expected if the elongated
galaxies were highly inclined disks.  It is at least as likely that 
the galaxies' elongated morphologies are due to merging subunits.

\end{abstract}


\section{Introduction}

At redshifts up to $z\sim1$, it is possible to
identify disk galaxies, place spectroscopic slits along the galaxies'
major axes, and
obtain rotation curves similar to those of disk galaxies in the local
universe (e.g.\ \citealt{vfp+96,vpf+97,bzs+03}).  As redshift increases
galaxy morphologies become increasingly irregular \citep{v01,cgj+03},
and spatially resolved
spectra for kinematic measurements are more difficult to obtain as
galaxy sizes approach the size of the seeing disk and the width of
slits.  In addition, the rest-frame optical emission lines used for such
measurements shift into the near-IR for $z\gtrsim1.4$.  Nevertheless
spatially resolved and tilted rest-frame optical emission lines from
which ``rotation curves'' can be constructed have been seen for galaxies
beyond $z\sim1.5$ \citep{pss+01,lcp+03,mvc+03,ess+03}; these
lines generally have low signal-to-noise ratios (S/N) and the
morphologies of the galaxies are irregular or unknown, making the
interpretation of the tilted lines uncertain.  One-dimensional
velocity dispersions which do not require spatial resolution are much
easier to obtain at high redshift; their use as a mass indicator
for such galaxies has been well-studied (e.g.\ \citealt*{kb00}), although
faint emission lines may not trace a galaxy's full gravitational
potential.  Whether 
or not an emission line is spatially resolved, the alignment of a slit with a
galaxy's apparent major axis should maximize the observed velocity,
if the galaxy is elongated because it is highly inclined.  Such a slit
orientation also tests the alignment of the morphological and kinematic
major axes. 

Observations such as these at $z\sim2$ are critical, as it has become
increasingly clear that this epoch is an 
important period in the 
evolution of the universe.  Galaxies at $z\sim3$ are compact, rapidly
star-forming, and morphogically disordered (e.g.\
\citealt{gsm96,ssa+01}), whereas those at 
$z \lesssim 1$ have become the normal Hubble sequence galaxies of the
universe today.  Recent results suggest that QSO activity reaches a
peak near $z\sim2$ \citep{fss+01}, and the median redshift of
bright sub-millimeter galaxies is $z=2.4$ \citep{cbis03}.
In addition, most of the stellar mass in the universe today formed
during this epoch: in a study of galaxies in the HDF-N,
\citet{dpfb03} find that 50--75\% of the mass in today's galaxies had 
formed by $z\sim1$, but only 3--14\% had formed by $z\approx2.7$.

In this paper we present kinematic measurements from $\Ha$ emission in
galaxies at
$z\sim2$ in the GOODS-N field, making use of the deep \textit{HST} ACS
imaging which has recently been obtained as part of the Great
Observatories Origins Deep Survey (GOODS; \citealt{goods03}).  We
describe our selection criteria, observations and data reduction in
\S\ref{sec:obs}, the results of our simple morphological analyses in
\S\ref{sec:morph}, and our kinematic results in \S\ref{sec:kin}.  In
\S\ref{sec:seeing} we highlight the critical importance of the seeing to
kinematic measurements at high redshift, and we discuss our results in
\S\ref{sec:disc}.  We use a cosmology with $H_0=70\;{\rm km}\;{\rm
  s}^{-1}\;{\rm Mpc}^{-1}$, $\Omega_m=0.3$, and $\Omega_{\Lambda}=0.7$
throughout.  In such a cosmology, 1\arcsec\ corresponds to 8.1 kpc at
$z=2.38$, the mean redshift of the current sample.

\begin{deluxetable*}{l l l c l l l l c l c}
\tablewidth{0pt}
\tabletypesize{\footnotesize}
\tablecaption{Galaxies Observed\label{tab:obs}}
\tablehead{
\colhead{Galaxy} & 
\colhead{R.A. (J2000)} &
\colhead{Dec. (J2000)} &
\colhead{Exposure time (s)} &
\colhead{$z_{abs}$\tablenotemark{a}} &
\colhead{$z_{\rm Ly\alpha}$\tablenotemark{b}} &
\colhead{$z_{\Ha}$\tablenotemark{c}} & 
\colhead{$\cal R$} &
\colhead{$G - \cal R$} & 
\colhead{$F_{\Ha}$\tablenotemark{d}} &
\colhead{Selection\tablenotemark{e}} 
}

\startdata

BX305 & 12:36:37.131 & 62:16:28.358 & $900 \times 4$ & 2.4825 & -- & 2.4839 & 24.28 & 0.79 & 4.2 & R\\
BX1055 & 12:35:59.594 & 62:13:07.504 & $900 \times 2$ & 2.4865 & 2.4959 & 2.4901 & 24.09 & 0.24 & 2.7 & E\\
BX1084 & 12:36:13.568 & 62:12:21.485 & $900 \times 5$ & 2.4392 & -- & 2.4403 & 23.24 & 0.26 & 7.3 & E\\
BX1085 & 12:36:13.331 & 62:12:16.310 & $900 \times 5$ & 2.2381 & -- & 2.2407 & 24.50 & 0.33 & 1.1 & T\\
BX1086 & 12:36:13.415 & 62:12:18.841 & $900 \times 5$ & -- & -- & 2.4435 & 24.64 & 0.41 & 1.8 & T\\
BX1277 & 12:37:18.595 & 62:09:55.536 & $900 \times 3$ & 2.2686 & -- & 2.2713 & 23.87 & 0.14 & 5.3 & E\\
BX1311 & 12:36:30.540 & 62:16:26.116 & $900 \times 4$ & 2.4804 & 2.4890 & 2.4843 & 23.29 & 0.21 & 8.0 & E\\
BX1322 & 12:37:06.538 & 62:12:24.938 & $900 \times 6$ & 2.4401 & 2.4491 & 2.4443 & 23.72 & 0.31 & 2.0 & E\\
BX1332 & 12:37:17.134 & 62:11:39.946 & $900 \times 3$ & 2.2113 & -- & 2.2136 & 23.64 & 0.32 & 4.4 & E\\
BX1368 & 12:36:48.241 & 62:15:56.237 & $900 \times 4$ & 2.4380 & 2.4455 & 2.4407 & 23.79 & 0.30 & 8.8 & R\\
BX1376 & 12:36:52.960 & 62:15:45.545 & $900 \times 4$ & 2.4266 & 2.4338 & 2.4294 & 24.48 & 0.01 & 2.2 & E\\
BX1397 & 12:37:04.115 & 62:15:09.837 & $900 \times 3$ & 2.1322 & -- & 2.1332 & 24.12 & 0.14 & 5.2 & E\\
BX1479 & 12:37:15.417 & 62:16:03.876 & $900 \times 5$ & 2.3726 & 2.3823 & 2.3745 & 24.39 & 0.16 & 2.5 & E\\

\enddata
\tablenotetext{a}{Vacuum heliocentric redshift from rest-frame UV
  interstellar absorption lines.}
\tablenotetext{b}{Vacuum heliocentric redshift of Ly$\alpha$ emission
  line, when present.}
\tablenotetext{c}{Vacuum heliocentric redshift of H$\alpha$ emission
  line.}
\tablenotetext{d}{Flux of H$\alpha$ emission line, in units of
  10$^{-17}$ ergs s$^{-1}$ cm$^{-2}$. This should be considered a
  lower limit because conditions were non-photometric.}
\tablenotetext{e}{Reason for selection.  E=elongated, T=``triplet,''
  additional galaxies on slit with BX1084, R=favorable redshift.  See
  \S2 for details.}

\end{deluxetable*}

\section{Observations and Data Reduction}
\label{sec:obs}

The 13 galaxies presented here lie in the Hubble Deep Field North
region imaged with the \textit{HST} Advanced Camera for Surveys (ACS)
as part of the GOODS program.  We have 
spectroscopically identified approximately 180 galaxies at $z\sim2$ in
this field, using $U_nG\cal R$ color selection criteria
and rest-frame UV spectra from the LRIS-B spectrograh
on the 10 m W.M. Keck I telescope on Mauna Kea \citep{ass+04,ssp+04}.  We have
observed a subset of 13 of these with the 
near-infrared spectrograph NIRSPEC \citep{mbb+98}, on the
W.M. Keck II telescope. 
The galaxies were selected primarily because they appeared
elongated on our initial inspection of the $z$-band GOODS-N
data (detailed morphological analysis was not done until the full V1.0
release of the GOODS data in late August 2003).  We also tried to
select galaxies with 
redshifts that put $\Ha$ in a favorable position with respect to the
night sky lines, and chose two (BX305, BX1368) for this reason, as
well as for their somewhat
more compact appearance for the sake of comparison.  BX1085 and BX1086
were observed because it was possible to place both of them on the
slit with the elongated galaxy BX1084, as discussed below.  The final column in
Table~\ref{tab:obs} summarizes the selection criterion for each
galaxy.

We attempted to align the slit with the elongated axis in all cases,
with the exception of BX1084, BX1085 and BX1086\footnote{BX1086 is the
only galaxy in the sample for which we did not previously know the
redshift.  We believe that the detected line is $\Ha$ because the
interloper fraction from the BX color selection criteria is $\sim9$\%,
and the typical interlopers (star-forming galaxies at $\langle z
\rangle = 0.17$) do not have a single strong emission line in
the $K$-band.  The $\Ha$ redshift is also nearly identical to the
redshift of BX1084, which is separated from BX1086 by 2\farcs9.}, which
lie on a single 
line and were therefore observed simultaneously with the PA set by
their relative orientation.  Coincidentally this PA
differs from the PA of BX1084 by only 19\degr.  Because morphological
analysis was done after the observations, the slit and galaxy PAs were
slightly misaligned, with an average offset of 10\degr.  11 out of
13 galaxies therefore had slits aligned to within 25\degr of the
galaxy PA, and 9
out of 13 to within 13\degr.  Such slight misalignments will not
prevent the detection of significant rotation, should it exist
\citep{vfp+96,sp98,bzs+03,mksp04}.  It is also possible for misalignment of
the slit and the galaxy to introduce the appearance of velocity shear
where none may actually exist; in \S~\ref{sec:seeing} below we explain
why this effect has not biased the current
observations.  The galaxy PAs and slit positions are
given in Table~\ref{tab:mk}.  

Most of the observations were conducted on the nights of 7-9 May 2003
(UT), with two additional objects (BX1055 and BX1397) observed on 7-8
July 2003 (UT).  We used the 0\farcs76 $\times$ 42\arcsec\ slit for all
observations.  The seeing was relatively poor during the May run, with
FWHM 0\farcs 7--0\farcs9 in the $K$-band; in July the seeing was better,
with FWHM $\sim$ 0\farcs 5.  Conditions were not photometric on either
run, and in particular much of the May data suffers from significant
losses due to cirrus clouds.  For a detailed description of the
observing and data reduction procedures, see \citet{ess+03}. 

\begin{figure*}[htbp]
\plotone{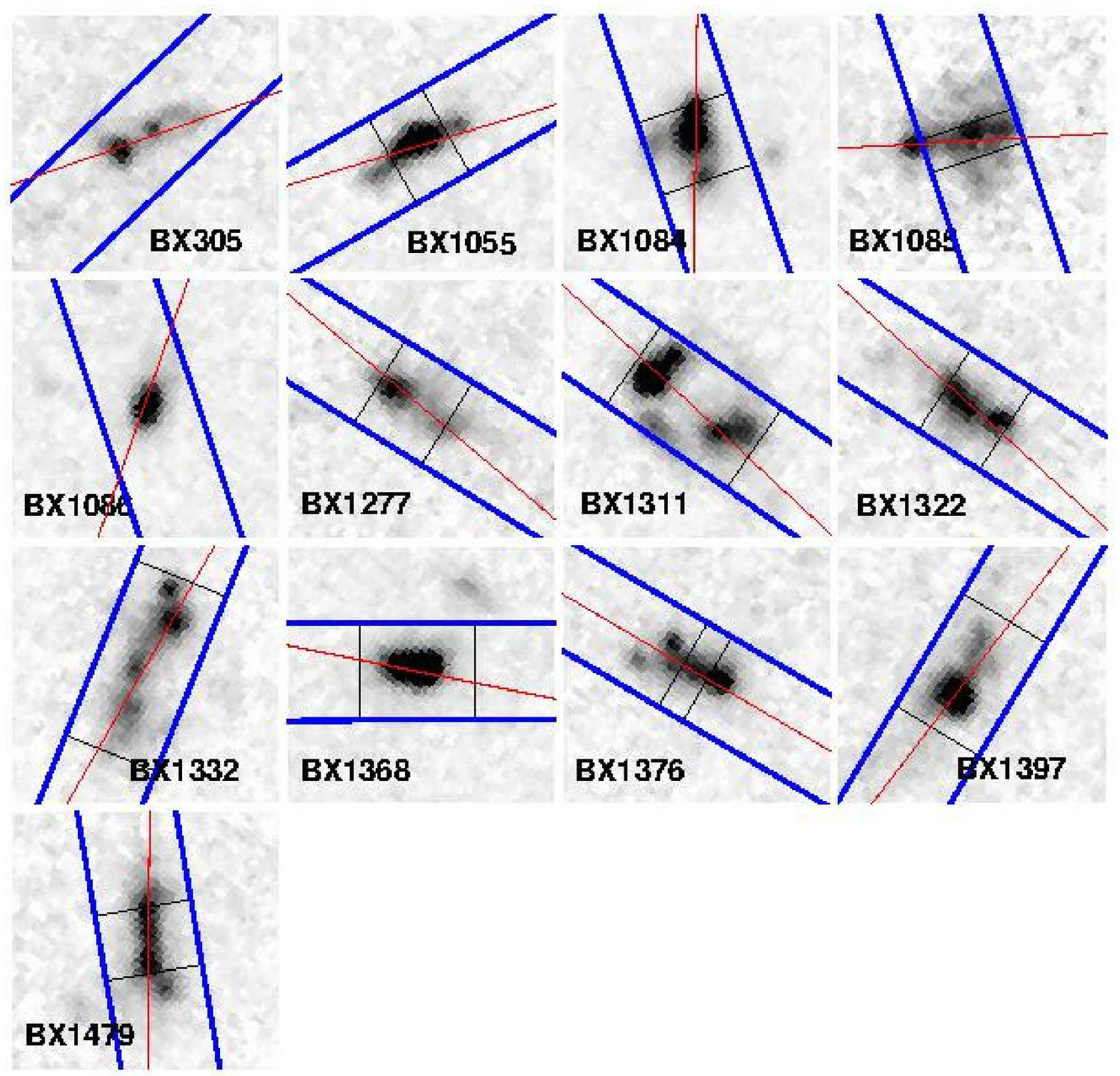}
\caption{\textit{HST} ACS images of the 13 galaxies in our sample,
  from the GOODS program.  The $BViz$ bands have been combined to
  maximize S/N, and the images shown cover the approximate range
  1300--2500 \AA\ in the rest frame.  The 0\farcs76-wide slit (6.2 kpc at
  $z=2.3$) is marked in blue,
  and the red line shows the galaxy PA.  The thin lines perpendicular
  to the slit show the approximate extent of the $\Ha$ emission after
  deconvolution of the seeing.  No lines are shown for BX305 and
  BX1086 because they appeared to be point sources after deconvolution.
  All images are oriented with 
  N up and E to the left, and the pixel scale is 0\farcs05/pixel.}
\label{fig:images}
\end{figure*}

\section{Morphologies}
\label{sec:morph}

\textit{HST} ACS 
images of the 13 galaxies in our sample are shown in
Figure~\ref{fig:images}.  For the purpose of morphological analysis we combined
the four ACS bands into a single image, weighting in order to maximize
the S/N.  The images shown therefore cover the approximate range
1300--2500 \AA\ in the rest frame.  The slits used are
marked with heavy blue lines and the object PA with fine red lines.
As noted in \S\ref{sec:obs}, the average 
misalignment of slit and galaxy is 10\degr; the slit and
galaxy PAs are aligned to within 25\degr\ for 11 of 13 objects, and to
within 13\degr\ for 9 of 13.  Most of the galaxies appear irregular on
simple inspection, and we have not attempted any kind of morphological
classification.  We have however performed some simple morphological analysis,
with two goals in mind: to determine whether there is any correlation
between a galaxy's kinematic properties and its
aspect ratio, and to determine a size to be used in mass estimates.

Morphological analysis of faint galaxies is difficult, as it requires
the separation of low surface brightness galaxy pixels from the sky
background.  We have estimated galaxy shapes and sizes using the
pixels that make up half of the sky-subtracted light within a 1\farcs5
(30 pixel)
radius around the object centroid.  The centroid is determined using an
iterative 
process that calculates the centroid of an object, subtracts the sky
value determined from a surrounding annulus,
and recomputes the centroid until convergence is reached.  We then use
the pixels within 1\farcs5 of
the final centroid for the remaining analysis.  We calculate two
measures of galaxy size as follows.  The effective 
half-light radius $r_{1/2} = (A/\pi)^{1/2}$, where $A$ is the area
of pixels encompassing 50\% of the galaxy's light, is sensitive
to how large the bright regions of an object are, but not to their
distribution; i.e.\ it depends on the number of pixels required to make up
half of the galaxy's light, but not on how those pixels are
distributed within the 30 pixel radius.  We find a mean and standard
deviation of $\langle 
r_{1/2} \rangle = 0\farcs24 \pm 0\farcs06$, corresponding to  $\langle 
r_{1/2} \rangle = 1.9 \pm 0.5$ kpc.  We 
also calculate $d_{\rm maj}$, the RMS dispersion of 
the light about the centroid along the major axis (we avoid using the
symbol $\sigma$ to eliminate confusion with velocity dispersions
calculated below), and the aspect ratio $a = d_{\rm min}/d_{\rm maj}$,
the ratio of the dispersions along the 
minor and major axes.  We find $\langle d_{\rm maj} \rangle = 0\farcs20
\pm 0\farcs09$, or $\langle d_{\rm maj} \rangle = 1.7 \pm 0.7$ kpc,
and  $\langle a \rangle = 0.41 \pm 0.12$.  We use  
$d_{\rm maj}$ for mass estimates.  Individual
values of these parameters are shown in Table~\ref{tab:mk}, and in the
following section we relate them to the galaxies' kinematic properties.

\section{Kinematic Results}
\label{sec:kin}

\begin{figure}[htbp]
\plotone{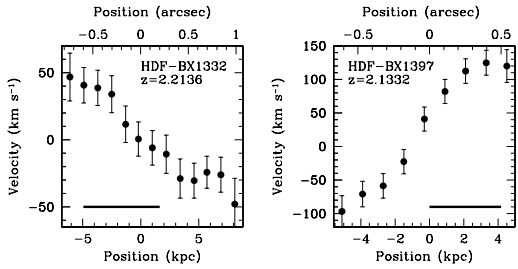}
\caption{Velocity curves for BX1332 and BX1397, the only two galaxies
  in the sample which show tilted emission lines.  The horizontal
  black lines represent the size of the seeing disk.  The points are
  strongly 
  correlated due to the seeing, but the spatial extent of each galaxy
  is larger than the seeing disk and the two endpoints used to
  determine $v_c$ are not correlated.}
\label{fig:newrot}
\end{figure}

Close examination of the 13 galaxies' two-dimensional spectra reveals
only two showing spatially resolved velocity shear.  We
show velocity curves for the two emission lines in
Figure~\ref{fig:newrot}, constructed by fitting a Gaussian profile in
wavelength at each spatial position along the emission line, summing
three pixels to increase the S/N.  We refer to these figures as
velocity curves rather than rotation curves because it is not
clear that the tilted lines are caused by rotation; see
\S\ref{sec:disc} for discussion of this issue.  The first of
the tilted emission lines, that of BX1332, extends almost 1\farcs75 in
the spatial 
direction; this is significantly larger than the $\sim0\farcs$8 seeing
disk, and therefore the modest tilt of $\sim90$ \kms\ 
peak-to-peak is readily 
apparent.  Assuming for the moment that this tilt is due to rotation,
we calculate a 
lower limit on the circular velocity $v_c$ of half of the total
velocity spread, $v_c \sim 45$ \kms.  We emphasize here that we use $v_c
\equiv (v_{max} - v_{min})/2$ as an observed quantity defined as half
of the velocity spread of the emission line; this is not the terminal
circular velocity, and is almost
certainly less than that velocity.  The second galaxy, BX1397, which
was observed in July 2003 when the seeing was 0\farcs5, has a smaller 
spatial extent of $\sim1.2$\arcsec\ but a larger velocity range, with
$v_c \sim 110$ \kms.  The large spatial
extent of BX1332 may be suggested by the fact that it is the most
elongated of the 13 galaxies, with $a=0.21$; but BX1397 has $a = 0.47$,
slightly higher than the mean $\langle a \rangle = 0.41$.  Clearly the
aspect ratio does not predict rotation. Such a low
incidence of 
rotation in galaxies with slits placed along their elongated axes is
perhaps surprising, especially since our earlier near-IR spectroscopy
found tilted lines in about 40\% of the galaxies observed, with random
(with one exception) slit orientations \citep{ess+03}.  In
\S\ref{sec:seeing} below 
we discuss limitations on observed rotational velocities imposed by the
seeing, and in \S\ref{sec:disc} we consider the implications of
our low incidence of observed rotation.  

It is also interesting to
compare the relative extent of the $\Ha$ and 
rest-frame UV emission.  BX1332 and BX1397 have values of $d_{\rm
  maj}$ of 0\farcs37 and 0\farcs20 respectively, but the total spatial
extent of all the pixels considered to be part of the galaxy for 
morphological analysis is $\sim1\farcs5$ for BX1332 and $\sim0\farcs9$
for BX1397.  Deconvolving the seeing from the $\Ha$ sizes above, we
find $\sim1\farcs6$ for BX1332 and $\sim1\farcs1$ for BX1397, values which
agree well with the total UV sizes.  We can make this comparison for
the rest of the objects by measuring the spatial width of the $\Ha$
emission line and again deconvolving the seeing.  The results of this
calculation are shown by the thin lines perpendicular to the slits in
Figure~\ref{fig:images}, which show the approximate extent of the $\Ha$
emission.  No lines are shown for BX305 and BX1086 because they
appeared to be point sources after deconvolution.  In general the
UV and $\Ha$ emission agree well, although there are
cases with greater extents of both UV (BX1376) and $\Ha$ (BX1368).
The spatial extent of the $\Ha$ emission line of each galaxy is given
in Table~\ref{tab:mk}.

We have calculated one-dimensional velocity dispersions for the
galaxies in the sample by fitting a Gaussian profile to each $\Ha$
emission line using the splot task in IRAF, which also provides
errors.  We deconvolved the instrumental profile by subtracting 
the instrumental FWHM of 15 \AA\ (measured
from the widths of sky lines) in quadrature from the Gaussian FWHM;
the instrumental profile is comparable to the widths of the lines, so
this deconvolution is important.  It is possible to obtain higher
resolution if the object and the seeing are smaller than the slit, but
because the seeing was comparable to the slit width for most of our
observations we have used the instrumental resolution as measured from
the sky lines.  After converting the FWHM to a velocity, we computed
the velocity dispersion $\sigma = \rm{FWHM}/2.355$. 
The results are shown in column 8 of Table~\ref{tab:mk}, with errors of
one standard deviation from propagating the error in the Gaussian line
fit.  Two of the lines (BX1055 and BX1322) had observed widths less
than 10\% greater than the instrumental profile, making the
deconvolution highly uncertain; for these lines we place an upper limit on
$\sigma$ of twice the one standard deviation error.  The measured width of
BX1085 was slightly less than the instrumental FWHM, preventing a
calculation of $\sigma$ entirely.  Neglecting the objects with upper
limits, we find a mean velocity dispersion of $\langle \sigma \rangle =
92$~\kms, with a standard deviation of 34 \kms; this is somewhat
smaller than the mean of the total sample of 61
galaxies at $z\sim2$ for which we currently have $\Ha$
measurements\footnote{These data will be discussed in full elsewhere.},
$\langle \sigma \rangle \sim114 \pm 51$ \kms.  We 
have combined these measurements of $\sigma$ with our estimates of galaxy
sizes  $d_{\rm maj}$ to determine masses.  We calculate the virial
mass $M_{\rm vir}=5\sigma^2 
(d_{\rm maj}/G)$, and find a mean and standard deviation of $\langle
M_{\rm vir} \rangle = (1.6 \pm 1.1) \times 10^{10} \msun$. 
The results of the mass calculations are shown in Table~\ref{tab:mk}.

\begin{figure}[htbp]
\plotone{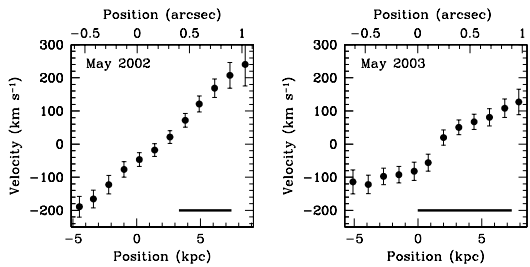}
\caption{Velocity curves for Q1700-BX691, observed in May 2002 when
  the seeing was $\sim0$\farcs5 (left) and in May 2003 when
  the seeing was $\sim0$\farcs9 (right).  The horizontal black lines
  represent the size of the seeing disk.
  The measured velocity amplitude decreases by a factor of $\sim2$
  when the galaxy is observed under poor conditions.}
\label{fig:bx691}
\end{figure}

\subsection{Seeing}
\label{sec:seeing}

\subsubsection{Tilted Emission Lines}

There is a long list of factors which affect the observed rotation
curves of disk galaxies; these include the alignment of the slit with
the major axis of the galaxy, the width of
the slit compared to the size of the galaxy, the inclination of the
disk, the spatial 
and velocity distribution of emission from the galaxy, and the blurring effect
of the seeing (see, e.g., \citealt{vfp+96,vpf+97,sp99,bzs+03,mksp04}
for discussion of these effects).  We 
concentrate here on the seeing, which can have a tremendous effect
on the velocities observed in galaxies at high redshift. 
We can see this in two different observations of the galaxy
Q1700-BX691 ($z=2.1895$), which we first observed with NIRSPEC in May 2002
\citep{ess+03}; during these observations the seeing was
$\sim0$\farcs5.  Using the identical instrumental configuration and
position angle on the sky, we observed the galaxy again in May 2003,
when the seeing was $\sim0$\farcs9.  In Figure~\ref{fig:bx691} we
compare velocity curves derived from the two observations: at left is
the previously published velocity curve, which has
$v_c \sim 220$ \kms, and on the right is a velocity curve from the
May 2003 observations, constructed in identical fashion and plotted on
the same scale for comparison.  It covers approximately the same
spatial range as 
the previous observation, but shows about half the range in velocity,
with $v_c \sim 120$ \kms.  It also shows some distortion in the shape of the
velocity gradient.  These changes are not unexpected: as the seeing
worsens, a wider range of velocities from different parts of the
galaxy is blurred together, and the emission from the core of the
galaxy, which presumably has both the highest surface brightness and
the lowest velocity, biases the velocity measurements out to larger
and larger radii.  Further, flux is lost as the seeing disk approaches
and exceeds
the size of the slit, the S/N declines as emission is spread over a
larger area, and poorer seeing makes it more difficult to center the
galaxy (or to be precise, the bright object we offset from) on the
slit.  

Q1700-BX691 (when observed under good conditions) shows the largest
velocity shear we have detected so far; in order to assess the
effects of the seeing on galaxies with a variety of velocity spreads,
we have smoothed their spectra in order to simulate poorer seeing.
This procedure was applied to all of the galaxies showing tilted lines
in the May 2002 
sample (with the exception of SSA22a-MD41, which was observed at
higher spectral resolution with the ISAAC spectrograph on the VLT).  We
smoothed by first replacing the emission line in the two-dimensional spectra 
with the sky background (copied from a spatially adjacent portion of
the slit, at the same wavelength) to make a sky frame, and then
subtracting this sky 
frame from the original image to create an image of the emission line
with the background removed.  This profile was then smoothed with a 
Gaussian filter in the spatial direction; to simulate seeing
of 0\farcs8 with a spectrum in which the original seeing was 0\farcs5,
we used a filter with 0\farcs6 FWHM ($0.5^2+0.6^2=0.8^2$).
In order to account for slit losses due to the
increase in seeing FWHM, we scaled the smoothed line by a factor of
$\sim0.6$, chosen empirically by measuring the decrease in line flux
in the two observations of Q1700-BX691 described above.  The smoothed
and scaled line was then added 
to the background frame to produce a final image.  We then measured a 
velocity curve as usual on the resulting two-dimensional 
spectra.  We find that the velocity spread is consistently reduced,
but the detailed results are somewhat unpredictable.  The effect of
the smoothing depends not only on the initial tilt of the line, but
also on the total flux and surface brightness.  Lines with initial
$v_c \lesssim 100$ \kms\ (Q1700-MD103, Q1623-BX511) no longer showed 
a tilt after smoothing, or had S/N too low to construct a velocity
curve; those with larger $v_c$ (West-BX600, Q1623-BX447) showed a
decrease in the velocity spread of approximately a factor of two, as
we saw in the case of Q1700-BX691.  The smoothed spectrum of
Q1700-BX691 itself has $v_c \sim 125$ \kms, a good approximation to the
May 2003 data which has $v_c \sim 120$ \kms.

We use these results to place upper limits on the internal velocity spread of
the ionized gas in the galaxies in
the current sample: the 11 galaxies with no detected shear, and 
BX1332 with $v_c \sim 45$ \kms, are likely to have $v_{0.5} < 100$
\kms, where we use the term $v_{0.5}$ to indicate the
velocity $(v_{max}-v_{min})/2$ that would be measured under
seeing conditions of $\sim0$\farcs5.  It
is important to note here that the one remaining galaxy, BX1397, which
shows the  
largest $v_c \sim110$ \kms, was observed in July 2003 when the
seeing was 0\farcs5 (as measured from a standard star observed
immediately before the spectra were taken).  Therefore all 13 of the
galaxies have $v_{0.5} \leq 110$ \kms.  In principle, of course, we
would like to know the intrinsic velocity $v_{0.0}$
rather than $v_{0.5}$; this has been estimated with detailed modeling of
rotation curves at lower redshift \citep{vfp+96,vpf+97}.  However,
such a calculation requires a specific galaxy model such as an
exponential disk, and 
knowledge of the galaxies' inclinations.  Given the irregular
morphologies of the galaxies in our sample, such assumptions are not
justified.  Even if we were to model an intrinsic
velocity distribution, it is likely that the results would be highly
degenerate; very different configurations of emitting material could
produce the same tilted line 
when significantly degraded by the seeing.  We therefore attempt to
estimate only $v_{0.5}$.

The seeing clearly reduces our ability to resolve the
velocity structure of a galaxy.  However, it also mitigates
the effects 
of slit misalignment, which under conditions of high spatial
resolution can introduce
spurious tilt in an emission line due to the correspondence between
position of the emission within the slit and wavelength in the
spectrum.  For example, if the slit width is 200~\kms\ and a
galaxy is tilted within the slit such that one end of it falls on one
side of the slit and the other end on the other side of the slit, it
will appear to have velocity shear of $\pm100$~\kms even if 
its actual velocity shear is zero.  This problem will arise only when
the FWHM of the galaxy light perpendicular to the slit is
significantly smaller than the width of the slit, a 
situation that is unlikely to occur even under our best seeing conditions of
$\sim0\farcs5$.  Given the slight slit misalignments of the current
observations, however, it is worthwhile to test the importance of this
effect.  We have done so by convolving the ACS image of
each galaxy with a Gaussian filter to approximate the seeing of the
NIRSPEC observation (under the assumption that the $\Ha$ emission
traces the UV continuum), and 
measuring the centroid of the galaxy within the slit at each spatial
position.  We find that the galaxy light fills the slit in all cases,
the maximum velocity introduced from the shift of the galaxy centroid
across the slit is $\pm15$~\kms, and the mean induced velocity is
$\pm9$~\kms; depending on the orientation of the galaxy, the velocity
shear we observe could be either increased or decreased by these
amounts.  These numbers are less than our typical errors in velocity. 
We have also tested the effect of 
error in the position of the slit by measuring the smoothed galaxies'
centroids within a simulated slit shifted by 0\farcs2 (our typical
uncertainty in positioning the galaxy on the slit) in either
direction perpendicular to the slit PA.  Again we find that the effect
is less than or comparable to our velocity errors, with a
maximum introduced velocity of $\pm21$~\kms\ and a mean of $\pm9$~\kms.
We therefore believe that only a small fraction of the velocity shear 
we observe in the tilted lines could be induced by misalignment of the slit.

A related concern involves the one-dimensional velocity dispersions.
As noted above and discussed further in \S\ref{sec:disc}, the
elongated galaxies for which we have placed slits on the extended axes
have a lower average velocity dispersion than the sample as a whole.
It is possible that by choosing random slit orientations we are
artificially elevating the velocity dispersion by illuminating the
slit unevenly as described above.  We have tested this by smoothing
the ACS images of known $z\sim2$ galaxies to approximate 0\farcs5
seeing, placing simulated slits along them at random orientations, and
measuring the shifts in the galaxies' centroids in the wavelength
direction at each spatial position along the slit.  For comparison
with the velocity dispersion $\sigma = \rm FWHM/2.355$, we define
$\sigma_{\rm induced} = \Delta v/2.355$, where $\Delta v$ is the full
velocity shift of the centroid.  We find that the
mean induced velocity dispersion $\langle \sigma_{\rm induced} \rangle= 8$
\kms, with a maximum of 21 \kms; 
these are nearly the same as the mean and maximum induced velocities
found above, suggesting that even with large errors in position angle
this effect does not introduce substantial velocity errors.  The effect of
$\sigma_{\rm induced}$ on the observed value of $\sigma$ will depend on the
source of the velocity dispersion: if the line widths are due to
random motions, then the induced velocity dispersion will add to the
true velocity dispersion in quadrature to produce the observed
profile, and $\Delta \sigma = \sigma_{\rm obs} - \sigma_{\rm true}$ is only a
few \kms.  On the other hand, if the velocity dispersions are due to
unresolved velocity shear, $\sigma_{\rm induced}$ could either increase or
reduce the observed $\sigma$ by up to $\sim20$ \kms, an amount that is
comparable to our typical error in $\sigma$.  The effect on the
average value of $\sigma$ should be quite small, however, since these
slit effects should increase or reduce $\sigma$ with equal
probability.  The true situation is likely some combination of these
two scenarios.  In either case, however, the use of random position angles is
unlikely to be a significant contaminant to our measurements of $\sigma$.

\subsubsection{Velocity Dispersions}

Although the seeing has a strong effect on the measurement of
spatially resolved 
velocity shear, it has a relatively small effect on the one-dimensional
velocity dispersion.  This is because spatial resolution is not required
to measure the velocity dispersion.  We can see this by considering
the limit in which the seeing disk exceeds the size of the galaxy.  In
this case we will be unable to distinguish emission from opposite
sides of the galaxy (the line will no longer be tilted), but emission
from both sides still contributes to the line width.  We will then still
measure the full velocity dispersion.  The situation is slightly more
complicated in practice, because of slit losses and decreased S/N: as
the seeing worsens emission from the edges of the galaxy may fall
outside the slit, decreasing the velocity dispersion somewhat, and S/N
degraded by the seeing may limit the detection of faint wings in the
line profile, especially if the galaxy's light is very centrally
concentrated.

We have quantified these effects with the observations and simulations
described above.  For Q1700-BX691 we measure $\sigma = 170 \pm 18$ 
\kms\ with 0\farcs5 seeing, and $\sigma = 156 \pm 29$ \kms\ with
0\farcs9 seeing; the two values of $\sigma$ agree within the errors.
We find that this is generally the case with artificially smoothed
data as well.  The line widths are usually well-preserved after the
smoothing, with $\gtrsim80$\% of the width retained; the weakest line
(Q1623-BX511)
shows a significant decrease, but also a 
substantial increase in the errors due to reduced S/N, such that the
two values are still within 1$\sigma$ of each other.  The S/N of this
degraded spectrum is lower than that of any object in the current
sample.  We therefore
believe that our one-dimensional measurements are relatively
uncompromised by the mediocre seeing conditions.  The fact that
the velocity spread of a tilted line is highly dependent on the seeing, 
while $\sigma$ is not, makes it very difficult to estimate
the former from measurements of the latter, as proposed
by \citet{ww03}. Thus, their conclusion that Lyman break galaxies are
preferentially low-mass star-bursting systems is premature.

\subsection{Large-scale motions}
\label{sec:winds}

\begin{figure}[htbp]
\plotone{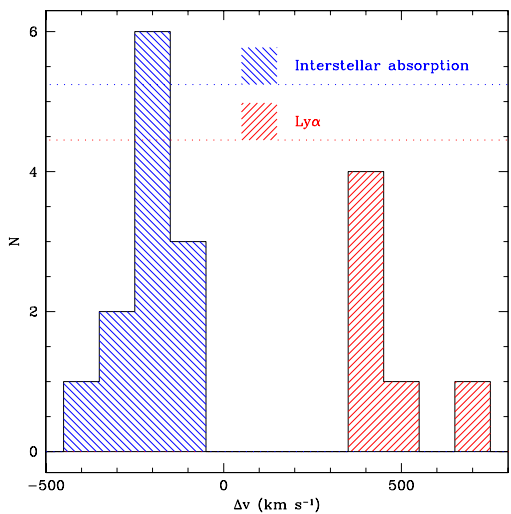}
\caption{Velocity differences between the UV interstellar absorption
  lines and $\Ha$ (blue histogram at left) and between $\lya$\ and $\Ha$
  (red histogram at right).  The offsets are consistent with
  galactic-scale superwinds with velocities of several hundred \kms.}
\label{fig:winds}
\end{figure}

Galactic-scale outflows, identified via the offsets between the
redshifts of the UV interstellar absorption lines or $\lya$ and the
nebular emission lines, are a common feature of galaxies at $z\sim3$
\citep{pss+01,ssp+03}.  These outflows typically have speeds of
$\sim300$ \kms\ with respect to the systemic redshift of the galaxy,
and are presumably powered by supernovae.  A comparison of the
redshifts given in Table~\ref{tab:obs} 
from the interstellar absorption lines, $\lya$, and $\Ha$ shows that a
similar pattern exists at $z\sim2$.  In Figure~\ref{fig:winds} we plot
histograms of the velocity offsets of the interstellar absorption
lines and $\lya$ with respect to $\Ha$; we see that the mean offset of
the absorption lines is $\langle \Delta v_{\rm abs} \rangle = -223 \pm 89$
\kms, and the 
mean offset of $\lya$ is $\langle \Delta v_{\lya} \rangle = 470 \pm 116$
\kms.  These velocities are consistent with those of all of the
$z\sim2$ galaxies for which we have performed this test
\citep{ssp+04}.  Note that these velocities are measured from the
centroids of the lines; we can estimate the terminal velocity of the
outflows by adding half of the interstellar absorption line widths
of $\sim650$ \kms\ \citep{ssp+04} to the numbers given above.

Could these outflows be the cause of any of the tilted emission lines we
have observed, or otherwise influence our kinematic measurements?
The outflow velocities ($\langle \Delta v_{\lya - \rm abs}
\rangle \sim 700$ \kms) are several times larger than either the
velocity spread of the tilted lines or the velocity dispersions, and
the results of \citet{ssp+03} show that nebular emission from HII
regions indicates a galaxy's systemic redshift rather than the
redshift of outflowing material: these authors
see both stellar photospheric lines and
nebular lines from HII regions in the composite spectrum of the
$z\sim3$ Lyman break galaxies, and find that their redshifts agree to
within 50 \kms.  \citet*{ham90} have shown that the emission line luminosity
of the outflow in M82 is $\sim10$\% of the luminosity of the total
galaxy.  \citet{chg+04} obtain a similar result in a study of the fraction
of the ISM ionized by non-radiative processes in four local starburst
galaxies, finding that 3--4\% of the $\Ha$ luminosity arises from such
shock-heated gas.  If these relations hold in more distant galaxies,
we would be 
unlikely to detect $\Ha$ from the outflows, especially given the
strong redshift dependence of surface brightness.  For these reasons
we believe that the observed outflows are unlikely to influence our
measurements of $\Ha$. 

\begin{deluxetable*}{l l r l l l l l l l l l l l}
\tablewidth{0pt}
\tabletypesize{\footnotesize}
\tablecaption{Morphologies and Kinematics\label{tab:mk}}
\tablehead{
\colhead{Galaxy} & 
\colhead{$z_{\Ha}$\tablenotemark{a}} & 
\colhead{PA\tablenotemark{b}} &
\colhead{Slit PA\tablenotemark{c}} & 
\colhead{$r_{1/2}$\tablenotemark{d}} & 
\colhead{$r_{1/2}$\tablenotemark{d}} &
\colhead{$d_{\rm maj}$\tablenotemark{e}} & 
\colhead{$d_{\rm maj}$\tablenotemark{e}} & 
\colhead{$a$\tablenotemark{f}} &
\colhead{$d_{\Ha}$\tablenotemark{g}} & 
\colhead{$d_{\Ha}$\tablenotemark{g}} & 
\colhead{$\sigma$\tablenotemark{h}} & 
\colhead{$v_c$\tablenotemark{i}} & 
\colhead{$M_{vir}$\tablenotemark{j}} \\
\colhead{} &
\colhead{} &
\colhead{} &
\colhead{} &
\colhead{(arcsec)} &
\colhead{(kpc)} &
\colhead{(arcsec)} &
\colhead{(kpc)} &
\colhead{} &
\colhead{(arcsec)} &
\colhead{(kpc)} &
\colhead{(\kms)} &
\colhead{(\kms)} &
\colhead{($10^{10}$ M$_{\odot}$)}
}

\startdata

BX305 & 2.4839 & 109.2 & \phs133.4 & 0.28 & 2.3 & 0.25 & 2.0 & 0.32 & -- & -- &     $140 \pm 31$ & -- & 4.1\\
BX1055 & 2.4901 & 107.0 & \phs120.0 & 0.14 & 1.2 & 0.09 & 0.7 & 0.52 & 0.48 & 3.8 &   $<124$ & -- & 1.2\\
BX1084 & 2.4403 & \phn$-0.6$ & \phs\phn18.7 & 0.24 & 1.9 & 0.16 & 1.3 & 0.58 & 0.59 & 4.8 &   $102 \pm 9$ & -- & 1.5\\
BX1085 & 2.2407 & \phn93.3 & \phs\phn18.7 & 0.29 & 2.4 & 0.25 & 2.0 & 0.39 & 0.27 & 2.2 &   -- & -- & --\\
BX1086 & 2.4435 & $-19.1$ & \phs\phn18.7 & 0.13 & 1.1 & 0.08 & 0.6 & 0.57 & -- & -- &     $67 \pm 42$ & -- & 0.25\\
BX1277 & 2.2713 & \phn49.8 & \phs\phn58.1 & 0.26 & 2.1 & 0.22 & 1.8 & 0.32 & 0.62 & 5.1 &   $63 \pm 11$ & -- & 0.83\\
BX1311 & 2.4843 & \phn47.2 & \phs\phn55.3 & 0.26 & 2.1 & 0.33 & 2.7 & 0.36 & 1.14 & 9.2 &   $88 \pm 13$ & -- & 2.3\\
BX1322 & 2.4443 & \phn47.7 & \phs\phn57.1 & 0.24 & 1.9 & 0.19 & 1.5 & 0.38 & 0.57 & 4.6 &   $<92$ & -- & 1.5\\
BX1332 & 2.2136 & $-29.9$ & \phn $-22.0$ & 0.32 & 2.6 & 0.37 & 3.1 & 0.21 & 1.47 & 12.1 &  $54 \pm 20$ & 45 & 1.0\\
BX1368 & 2.4407 & \phn78.8 &\phs\phn 90.3 & 0.15 & 1.2 & 0.09 & 0.7 & 0.54 & 0.90 & 7.3 &   $138 \pm 12$ & -- & 1.5\\
BX1376 & 2.4294 & \phn59.7 & \phs\phn59.6 & 0.24 & 2.0 & 0.20 & 1.6 & 0.37 & 0.22 & 1.8 &   $96 \pm 31$ & -- & 1.8\\
BX1397 & 2.1332 & $-37.4$ & \phn $-31.0$ & 0.30 & 2.5 & 0.20 & 1.6 & 0.47 & 1.02 & 8.4 &   $123 \pm 22$ & 110 & 2.8\\
BX1479 & 2.3745 & \phn$-0.5$ & \phs\phn\phn9.7 & 0.22 & 1.8 & 0.23 & 1.9 & 0.24 & 0.51 & 4.2 &    $46 \pm 28$ & -- & 0.45\\

\enddata
\tablenotetext{a}{Vacuum heliocentric redshift of H$\alpha$ emission
  line.}
\tablenotetext{b}{Position angle of galaxy in degrees.}
\tablenotetext{c}{Position angle of slit in degrees.}
\tablenotetext{d}{Effective half-light radius; see
  \S\ref{sec:morph}.}
\tablenotetext{e}{RMS dispersion of light about centroid along major
  axis; see \S\ref{sec:morph}.}
\tablenotetext{f}{Aspect ratio of RMS dispersions along minor and major axes;
  see \S\ref{sec:morph}.} 
\tablenotetext{g}{Approximate spatial extent of $\Ha$ emission line.}
\tablenotetext{h}{One-dimensional velocity dispersion from $\Ha$
  emission line.}
\tablenotetext{i}{($v_{max} - v_{min})/2$ for those galaxies which
  show tilted emission lines.}
\tablenotetext{j}{Virial masses calculated from $\sigma$ and $d_{\rm
    maj}$; see \S\S\ref{sec:morph} and \ref{sec:kin}.}
\end{deluxetable*}

\section{Discussion}
\label{sec:disc}

We have observed a lack of evidence for rotation in the form of tilted
emission lines in elongated 
galaxies at $z\sim2$, in spite of the fact that our spectroscopic
slit was nearly aligned with the galaxies' major axes in 11 out of 13
cases.  The seeing 
was mediocre during many of our observations, and we show with 
observations of the same object taken under different conditions and with
artificially smoothed data that this can affect the observed
velocity shear at a given radius by as much as a factor of two. 
Accounting for the effects of the seeing, we believe that all 13 of
the galaxies in the present sample have $v_{0.5} \leq 110$ \kms; 11 of them
show no evidence of rotation at all.  By $v_{0.5}$ we
mean half of the peak-to-peak velocity amplitude measured in an
exposure of $\sim$1 hour under good seeing conditions
($\sim0$\farcs5); unlike the
terminal velocities of rotation curves of local galaxies, this is not
a fundamental quantity, and it likely underestimates the true 
velocity.  This observed lack of rotation is perhaps surprising;
rotational velocities of a few hundred \kms\ are predicted for models
of star-forming galaxies at $z\sim2$--3 \citep*{mmw98,mmw99}, and we
know that galaxies with $v_c \sim 200$ \kms\ are present among the
still relatively small sample with rest-frame optical emission line
spectra at these redshifts \citep{lcp+03,ess+03}.

How, then, are we to interpret the lack of observed rotation in
galaxies with slits aligned with their apparent major axes?  The kinematic
measurements indicate that most of the galaxies are not disks with irregular
morphologies due to knotty star formation.  In fact we find
that the alignment of the slit with a galaxy's major axis has little effect
on whether or not a tilted emission line is observed.  Of our total
sample of 29 galaxies (the 13 presented here and 
the 16 in \citealt{ess+03}), 12 had slits aligned with their major axes
and 17 did not.  We see tilted emission lines in 25\% of the aligned
galaxies and 29\% of the unaligned galaxies; it seems that we are
slightly more likely to observe a tilted line by choosing a random
position angle, although this is hardly a robust result given the small
numbers involved and the effect of the seeing on the detectability of
velocity shear.  

Given these results, we should consider the possibility
that the tilted emission lines, when observed, may not always be due
to rotation.  Two or more merging clumps of material with
relative velocities of $\sim100$ \kms, blurred by the seeing, could
produce an extended, tilted emission line.  Such a scenario fits
naturally within the context of hierarchical structure formation, and
has been used to explain the profiles of damped $\lya$\ absorption
systems at high redshift \citep*{hsr98}.  In this model the
measured velocity would depend on the orientation of the
merging clumps, and projection effects would be difficult to quantify.  This
is not meant to be an exclusive explanation for the origin of the
tilted lines; rotation could certainly still be a factor as well,
especially in the case of objects such as Q1700-BX691 that show a 
regular pattern of high velocity shear over a large spatial area.

\begin{figure}[htbp]
\plotone{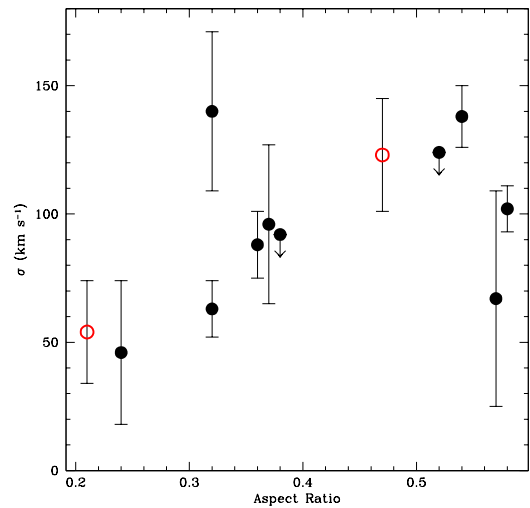}
\caption{The velocity dispersion $\sigma$ plotted against the aspect
  ratio $d_{min}/d_{maj}$.  The two open red symbols are those
  galaxies with tilted 
  emission lines.  We see mild evidence that the more symmetric
  galaxies have larger velocity dispersions.}
\label{fig:a-sigma}
\end{figure}

The uncertain causes and unreliable velocities of the tilted lines make
inferences from this type of kinematic measurement difficult, and it
therefore makes sense to focus our attention on what can be learned
from the one-dimensional measurements, which are relatively unaffected by the
seeing.  In our simulations at least 80\%\ of the velocity width was
retained after smoothing in all except one case; after the smoothing,
the exception had S/N lower than than of any object in the current
sample.  We have therefore compared the velocity dispersions of the
sample with their morphologies; we see some evidence that the more elongated
galaxies may have smaller velocity dispersions than 
the sample as a whole.  Considering only those galaxies which we
selected because they appeared elongated (i.e.\ neglecting BX305,
BX1085, BX1086 and BX1368), we find a mean velocity dispersion of 80
\kms, with an error in the mean of 9 \kms.  We compare this with the
mean velocity dispersion of the remaining 54 galaxies at $z\sim2$ for which we
have $\Ha$ measurements (these data will be discussed in full
elsewhere), $\langle \sigma \rangle = 118 \pm 7$ \kms; the difference
in the means is more than 3$\sigma$.  Using a K-S test, we find that the
probability that the two samples are drawn from the same population is
0.04.  To investigate this further we plot in
Figure~\ref{fig:a-sigma} 
the aspect ratio $a = d_{min}/d_{maj}$ against the velocity dispersion
$\sigma$.  There is mild evidence for a correlation, in the sense that
the more symmetric 
galaxies also have larger velocity dispersions; with a Spearman rank-order
correlation test we find that the probability of observing these data
if there is no correlation is 0.13.  

These nine galaxies have $\langle a \rangle = 0.38$, with an error in
the mean of 0.04.  We compare this with the 90 other galaxies we have
identified at $2.0 < z <  2.7$ in the GOODS-N field. These have
$\langle a \rangle = 0.51$, with an error in the mean of 0.02; the
difference in the means is $3\sigma$, showing that this subsample is
significantly more elongated than one made up of galaxies chosen at
random.  We have also calculated the mean 
$E(B-V)$ (as determined from the $G-{\cal R}$ colors, assuming a
star formation age of $>10^7$ yrs and a Salpeter IMF) of the elongated
sample, $\langle E(B-V) \rangle = 0.105\pm0.017$ (we again quote the
mean and the error in the mean).  As above we compare this with the 
remaining galaxies with $2.0<z<2.7$ in the GOODS-N
field, which have $\langle E(B-V) \rangle = 0.149\pm0.009$, and find
that the
elongated galaxies have a lower $\langle E(B-V) \rangle$ by 2$\sigma$.
These nine galaxies are also somewhat brighter than the sample as a
whole, with $\langle{\cal R}\rangle=23.87 \pm 0.15$ as compared to
$\langle{\cal R}\rangle=24.32 \pm 0.06$\footnote{Because our only
  photometry is in the rest-frame UV, we defer the calculation of
  absolute magnitudes until we have obtained rest-frame optical
  photometry.}; we see no correlation between $\cal R$ magnitude and
$E(B-V)$, however. 
Both the decreased reddening and the lower velocity dispersions are
the opposite of what one would expect if these galaxies were
highly inclined disks:  increasing the inclination of a disk also
makes it appear redder, and
the alignment of the slit and the major axis should maximize the
observed velocity.

We have seen that elongated galaxies sometimes, but not always, show evidence
for velocity shear along their major axes, and that clumpy galaxies that
appear merger-like sometimes, but not
always, show evidence of shear as well.  We also see shear in galaxies
of unknown morphology with the slit misaligned by an unknown amount.  The
situation is clearly complicated, and it is unlikely that a single model
will explain all of these results; some combination of rotation and
merging seems to be the most likely answer.  It will be
challenging to distinguish between these possibilities with 
the type of measurements presented here, however.  Even under ideal conditions,
the blurring of the seeing makes it difficult to discriminate between
rotation and merging.  Evidence for rotation could be found in deeper
observations if the velocity curves are seen to flatten like the
rotation curves of
local galaxies, and more detailed morphological analysis may help pick
out galaxies that are likely to be mergers.  The best way to
distinguish between the proposed scenarios, 
however, would be to obtain spectroscopic observations of an entire
galaxy at high spatial resolution, and we look forward to observations
of objects such as these with the new generation of near-IR integral
field spectrographs with adaptive optics. 

\acknowledgements
We thank the anonymous referee for useful comments.  We would also like
to thank the staff at the Keck Observatory for their 
competent assistance with our observations.  CCS, DKE, and AES have been
supported by grants
AST00-70773 and AST03-07263 from the U.S. National Science Foundation
and by the David 
and Lucile Packard Foundation.  AES acknowledges support from the
Miller Institute for Basic 
Research in Science.  Finally, we wish to extend special thanks
to those of Hawaiian ancestry on whose sacred mountain we are privileged
to be guests.  Without their generous hospitality, the
observations presented herein would not have been possible.



\clearpage

\end{document}